\documentclass[%
 preprint,%
 amssymb, amsmath,%
 prl,cha,superscriptaddress,%
]{revtex4-1}
\usepackage{graphicx}
\usepackage{comment}
\usepackage[capbesideposition=right]{floatrow}
\usepackage{sidecap}
\usepackage{nicefrac}
\usepackage{docs}%
\usepackage{bm}%
\usepackage[colorlinks=true,linkcolor=blue]{hyperref}%
\expandafter\ifx\csname package@font\endcsname\relax\else
 \expandafter\expandafter
 \expandafter\usepackage
 \expandafter\expandafter
 \expandafter{\csname package@font\endcsname}%
\fi
\begin{document}
	
\title{Domain Wall Spin Structures in Mesoscopic Fe Rings probed by High Resolution SEMPA}%

\author{Pascal Krautscheid}%

\affiliation{Institut f\"ur Physik, Johannes Gutenberg-Universit\"at Mainz, Staudinger Weg 7, 55128 Mainz, Germany}%
\affiliation{Graduate School of Excellence Materials Science in Mainz, Staudinger Weg 9, 55128 Mainz, Germany}%

\author{Robert M. Reeve}%

\affiliation{Institut f\"ur Physik, Johannes Gutenberg-Universit\"at Mainz, Staudinger Weg 7, 55128 Mainz, Germany}%

\author{Maike Lauf}%

\affiliation{Institut f\"ur Physik, Johannes Gutenberg-Universit\"at Mainz, Staudinger Weg 7, 55128 Mainz, Germany}%

\author{Benjamin Kr\"uger}%
\affiliation{Institut f\"ur Physik, Johannes Gutenberg-Universit\"at Mainz, Staudinger Weg 7, 55128 Mainz, Germany}%

\author{Mathias Kl\"aui}
\email{klaeui@uni-mainz.de}
\affiliation{Institut f\"ur Physik, Johannes Gutenberg-Universit\"at Mainz, Staudinger Weg 7, 55128 Mainz, Germany}%
\affiliation{Graduate School of Excellence Materials Science in Mainz, Staudinger Weg 9, 55128 Mainz, Germany}%

\date{2016}%

\begin{abstract}

We present a combined theoretical and experimental study of the energetic stability and accessibility of different domain wall spin configurations in mesoscopic magnetic iron rings. The evolution is investigated as a function of the width and thickness in a regime of relevance to devices, while Fe is chosen as a material due to its simple growth in combination with attractive magnetic properties including high saturation magnetization and low intrinsic anisotropy.
Micromagnetic simulations are performed to predict the lowest energy states of the domain walls, which can be either the transverse or vortex wall spin structure, in good agreement with analytical models, with further simulations revealing the expected low temperature configurations observable on relaxation of the magnetic structure from saturation in an external field. In the latter case, following the domain wall nucleation process, transverse domain walls are found at larger widths and thicknesses than would be expected by just comparing the competing energy terms demonstrating the importance of metastability of the states. The simulations are compared to high spatial resolution experimental images of the magnetization using scanning electron microscopy with polarization analysis to provide a phase diagram of the various spin configurations. In addition to the vortex and simple symmetric transverse domain wall, a significant range of geometries are found to exhibit highly asymmetric transverse domain walls with properties distinct from the symmetric transverse wall. Simulations of the asymmetric walls reveal an evolution of the domain wall tilting angle with ring thickness which can be understood from the thickness dependencies of the contributing energy terms. Analysis of all the data reveals that in addition to the geometry, the influence of materials properties, defects and thermal activation all need to be taken into account in order to understand and reliably control the experimentally accessible states, as needed for devices.

\end{abstract}
\maketitle
\section{Introduction}
The ability to control the magnetic spin configuration of nanoscale magnetic structures is a vital initial requirement for spintronic devices~\cite{spintroncDevices}. The materials properties of a system are one key factor and hence varying the material of a device is one handle to control the magnetization state. Other dominating contributions in patterned thin film structures are shape and configurational anisotropy, since there is an energetic advantage for the system to reduce the generated magnetic stray field by forming flux closure states in order to avoid having magnetization directed perpendicular to the sample edges~\cite{hubert, nanoshape}. Hence geometrical device design has become possible with the advent of high quality thin film deposition combined with nanoscale lithography to achieve robust,  tailored spin configurations. One particularly convenient and well-studied geometry is that of thin-film mesoscopic magnetic rings which exhibit particularly simple and robust spin-configurations. Such ring structures have received interest for potential applications such as MRAM elements~\cite{ringMRAM} or magnetic logic~\cite{ringlogic, ringlogic2}. Provided the ring is narrow enough, it is energetically preferable for the magnetization to track the edge of the structure, leading to the lowest energy configuration being the quasi-uniform flux-closure state which is termed the vortex state~\cite{vortexformation}. On relaxation from saturation the metastable so-called onion state can result, which is characterized by two magnetic domain walls on opposing sides of the ring~\cite{onionvortex, ringMFM, ringLorentz, MKheadtohead}. \newline For devices, the detailed spin structure of a domain wall is crucial for setting the relevant physical properties such as the dynamic behaviour including domain wall velocities and critical current densities for current induced domain wall motion~\cite{MKheadtohead, Boulle2011159, refinedphasepy, DWmobility, ringveltype} and hence this can determine the ultimate performance and attainable data storage densities. Two types of domain wall are favored in such nanoscale planar wires; in general, in narrower and thinner structures, the transverse domain wall is observed where the magnetization rotates by 180$^\circ$ via a roughly triangular region where the magnetization is directed off-axis to the wire~\cite{McMichaelDonahue,Boulle2011159, MKheadtohead,refinedphasepy}. Conversely, for wider and thicker structures, the so-called vortex wall is preferred where the magnetization curls around a central vortex core region in which the magnetization is directed out of the plane of the structure~\cite{McMichaelDonahue,Boulle2011159, MKheadtohead,refinedphasepy}. The energetically favoured state is determined by the interplay between dipolar and exchange energy contributions which scale differently with the geometry. Beyond such qualitative considerations, for the workhorse system Permalloy, the quantitative details of the phase diagram are well known~\cite{thermalDW, refinedphasepy}, however, the materials’ parameters of the system also influence the range of sizes where the different domain wall types are stable as has been shown for the case of cobalt~\cite{cowallphase}. 

For spintronic devices based on domain wall motion, materials with a large spin polarization are desirable for large torques~\cite{thiaville2005}, while additionally low intrinsic magnetic anisotropies are required in order to reduce pinning effects and to provide a robust system where the magnetization configurations can be tailored via the geometry. Fe could be an attractive material in this regard, since it has a lower magnetocrystalline anisotropy than Co, while retaining a high spin polarization and the highest saturation magnetization of the elemental ferromagnets. Furthermore, while some recent research interest has shifted to more exotic materials such as highly spin polarized Heusler alloys and oxides~\cite{finizio}, there remain barriers to the industrial adoption of such systems due to the difficulty in obtaining reliable growth conditions on industrially relevant substrates for large scale production. Hence, there remains great potential for use of simple materials in real applications, however surprisingly in the case of iron an in-depth characterization of the domain wall spin structures is so far lacking, despite the aforementioned advantageous magnetic properties of this material. While many studies just consider the two types of domain wall mentioned above, it has been predicted~\cite{refinedphasepy} and experimentally confirmed~\cite{holoDW} that the transverse domain wall can occur in both symmetric and asymmetric configurations, which are expected to have different properties. For example, it has recently been revealed that the symmetry of the transverse domain wall is important in determining the depinning process from domain wall traps~\cite{asymTDWdepin} and wire kinks~\cite{DWKinks}. Additionally in the presence of transverse applied fields, asymmetric transverse domain walls were observed to exhibit new dynamic behaviours with the upper and lower edges of the wall propagating at different velocities during longitudinal field driven motion, leading to a gradual increase in the wall asymmetry~\cite{DWsplitting}. This means that it is vital to also investigate the degree of asymmetry of the transverse domain walls with geometry, which has been previously neglected for most studies and requires suitably high-resolution magnetic imaging to discriminate the sometimes subtle differences between the wall types.

In this work we study the occurrence of the three mentioned domain wall types while varying the dimensions of iron rings using both micromagnetic simulations and direct high resolution magnetic imaging via scanning electron microscopy with polarization analysis (SEMPA)~\cite{LSMO_SEMPA, SEMPArev}. Within the simulations we systematically change the width and thickness of the structures and determine the resulting domain wall type with the lowest energy. The simulations provide a good qualitative understanding of the competing energy terms and show excellent agreement with an analytical model of the system, yet do not provide a good quantitative agreement with the experimentally observed wall types. For a fuller understanding we perform further simulations which include the process of domain wall nucleation on relaxing the spin structure from the initial state. A phase diagram of the different wall types is presented which we show can only be fully understood by taking into account the experimental domain wall initialization procedure in addition to the effects of thermal activation and sample defects.

\section{Experimental and Numerics}
\label{sec:exp}
In order to calculate the expected energetic stability of the different domain configurations in iron we performed micromagnetic simulations using the MicroMagnum code~\cite{micromag}. The Landau-Lifshitz-Gilbert equation describes the dynamic behaviour of the magnetization, $\mathbf{M}$, as follows~\cite{LLG1}:
\begin{equation}\frac{d\mathbf{M}}{dt}=-\gamma \mathbf{M} \times \mathbf{H_{eff}}-\alpha  \mathbf{M} \times \frac{d\mathbf{M}}{dt}\label{eqn:LLG},\end{equation} 
where $t$ is the time, $\gamma$ is the gyromagnetic ratio, $\alpha$ the Gilbert damping~\cite{LLG2} and $\mathbf{H_{eff}}$ the effective field. This equation is solved numerically via a finite difference method for the energy minimization procedure which is iterated until convergence. In contrast to the frequently studied straight wires, in this work curved structures are investigated which have particular merits in both experiment and applications. Such geometries facilitate the nucleation of domain walls at desired angular locations by the simple relaxation of the magnetization from a saturated state along a desired direction~\cite{MKheadtohead} and do not require the injection of domain walls from adjacent nucleation pad regions, which can be prone to stochasticity concerning the obtained spin structures within the wire. Half rings were simulated on an $800\times400\times1$ grid with the rectangular cuboidal cells having in-plane dimensions of 1.25\,nm and a thickness given by $d$, the thickness of the film being simulated. For thin films it is known that the magnetization is essentially uniform through the depth of the film and hence having just a single cell for the z direction is a reasonable assumption which helps to reduce the required simulation times.\newline
The simulations in this work were performed for a range of half ring sizes for widths between $w=$ 30 and 400\,nm and thicknesses between $d=$ 5 and 40\,nm. In order to reduce the computation time the outer diameter (O.D.) was kept at 1\,$\mu$m for all widths and instead of the experimentally studied full ring, a half-ring shaped wire was simulated which does not change the results as checked for a few selected geometries. Standard materials parameters for iron were chosen equivalent to $M_{s}=1.7\times10^{6}$\,A/m and $A=2.1\times10^{-11}$\,J/m~\cite{oomfFE}. For each geometry two simulations were initially performed to find the lowest energy state, one starting from a vortex wall configuration and the other starting from a transverse wall. The states were then relaxed and the total energy of the final configuration calculated in each case to compare the relative stability. In a first stage of relaxation a small external field was applied along the y axis of $\mu_0 H \approx$ 2\,mT, in order to stabilize the position of the domain wall in the centre of the half-ring and prevent it from migrating to the half ring ends where it can be expelled, resulting in a uniformly magnetized state. In real systems naturally occurring defects play an equivalent role and act as small pinning centres for the walls, providing an energy barrier between the metastable onion state and the lowest energy vortex state. In a second stage of relaxation the field is removed and the relaxation was then continued until the state was suitably converged, defined as a rate of change of magnetization of less than 0.01$^\circ$ per ns. 

In order to experimentally investigate the actually occurring domain wall types magnetic imaging of the domain wall configurations in iron rings was performed. For the imaging of the domain wall spin configurations in rings of other materials, previous work has employed electron holography~\cite{holoDW} or Lorentz microscopy, which require that the samples are fabricated on delicate membranes for the transmission measurements~\cite{ringLorentz}, photo-emission electron microscopy~\cite{widerings, thermalDW, cowallphase}, which is mainly available at large-scale facilities and can be limited in its resolution, or magnetic force microscopy (MFM)~\cite{ringMFM, thirdwall, vortexheadtohead,hayashi}, which can modify the spin configuration of the sample and is however sensitive only to the stray magnetic field from a sample and therefore harder to relate directly to the spin structures obtained from simulations. Therefore, in this paper we chose the imaging technique SEMPA~\cite{SEMPArev}, which is a powerful lab-based method with an excellent spatial resolution of less than 20\,nm and which can provide quantitative direct information concerning the spin configurations~\cite{SEMPArings}. Further details of the setup are provided in Ref.~\cite{LSMO_SEMPA}.

The samples were fabricated on Si/SiO$_2$ substrates via a standard electron beam lithography procedure followed by lift-off. Iron was deposited by ultra-high vacuum thermal evaporation at rates around 4\,nm/hour up to the desired thickness and a $\sim$ 2.5\,nm gold capping layer was employed, as needed, to counter oxidation of the magnetic material. In the case of the 20\,nm thick rings we started from the measured 24.5\,nm samples and reduced the thickness using 1\,kV in-situ argon ion sputtering. For the thicker samples between 15 and 26\,nm in thickness we chose rings of 2\,$\mu$m O.D., however for the thinner structures in the vicinity of the expected phase transition (thickness 10\,nm $\ge d \ge$ 5\,nm) we chose rings of a fixed outer diameter of 1\,$\mu$m in order to provide a more direct comparison with the simulations. Different ring widths were prepared between 90 and 310\,nm for the 1\,$\mu$m O.D. rings or between 100 and 750\,nm in the case of the 2\,$\mu$m O.D. structures by changing the inner diameter. Whilst the simulations modelled isolated half ring structures the experimental structures are arrays of full rings each containing  two domain walls following initialization. As such, in the experiments we also need to consider the possibility of stray field interaction between adjacent domain walls both within and between rings, which can potentially lead to spacing-dependent transitions between domain wall types~\cite{DWintenergy, DWintPlanar}. In order to rule out such effects, neighbouring rings were separated by more than the ring diameter~\cite{ringinteract}. Furthermore the maximum studied ring width was limited to $\sim350$\,nm for the 1\,$\mu$m O.D. in order to have a separation of several hundred nm between the two walls in the same ring, which avoids significant coupling effects taking into account a slightly higher stray field interaction for domain walls in Fe as compared to that previously measured for Co due to the difference in magnetostatic energies ($\sim M_s^2$)~\cite{DWintenergy}. Before imaging, the surface of the samples was cleaned using short argon ion milling in order to remove any gold capping layer and/or oxide from the surface, which is necessary due to the extreme surface sensitivity of SEMPA. The magnetic configuration of the samples was then initialized by applying a saturating magnetic field and relaxing the state, in order to generate the onion state with two domain walls in the ring. Imaging was then performed at ambient temperature.

\section{Results}
\label{sec:res}
\begin{figure}[b!]
\centering
\includegraphics[width=1\textwidth,keepaspectratio=true]{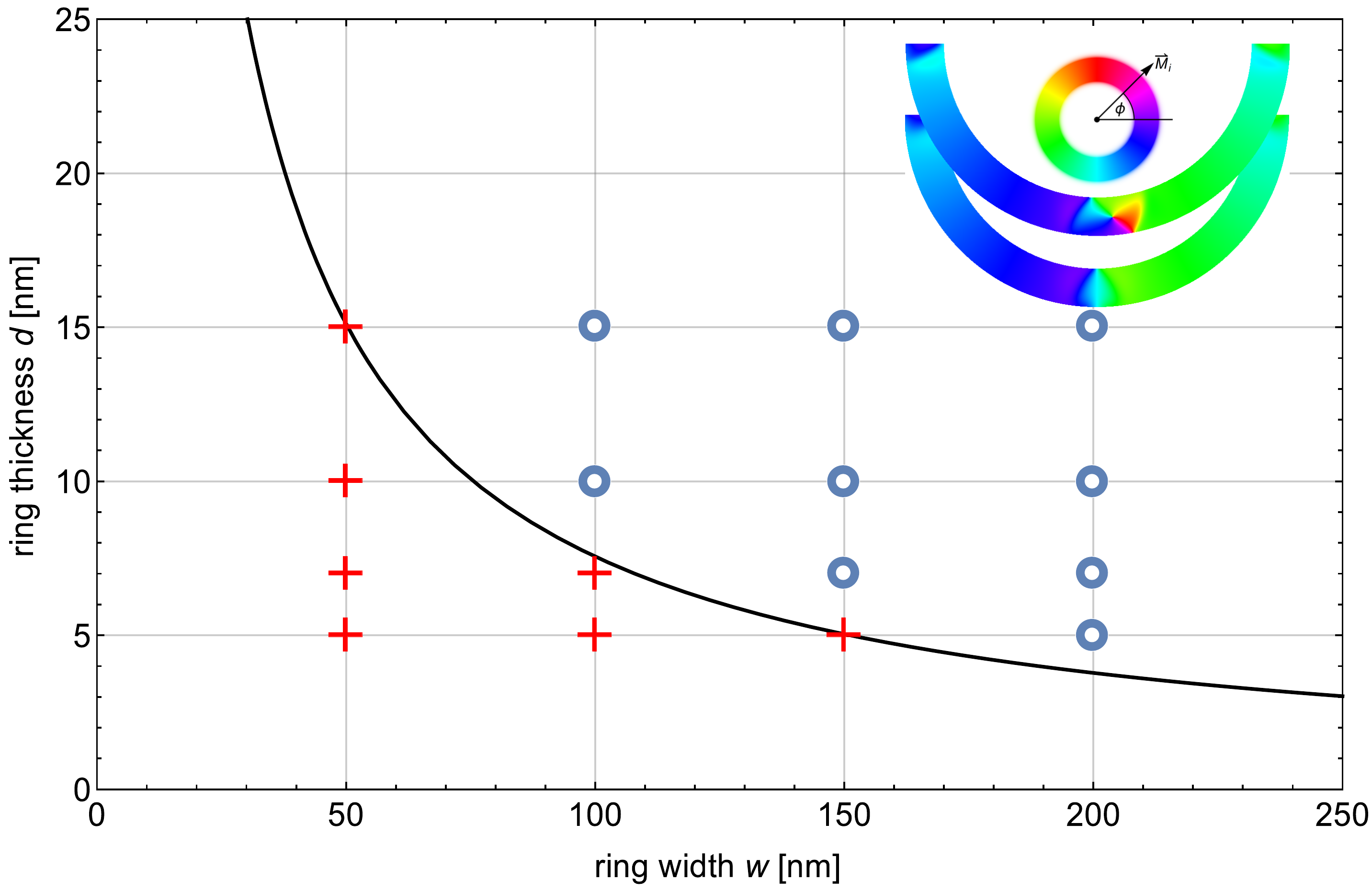}
\caption{Phase diagram of the lowest energy domain wall configurations for different ring sizes from micromagnetic simulations (Using MicroMagnum~\cite{micromag}). A red cross denotes transverse walls and a blue circle vortex walls. The line is a fit to the data, with Equation~\ref{eqn:phase} using $C \times \delta^2 = 756 \pm 17\,nm^2$, following McMichael and Donahue~\cite{McMichaelDonahue}. The inset depicts simulated head-to-head domain wall configurations in iron half-rings of thickness 25\,nm (top) and 7\,nm (bottom), with an outer diameter of 1\,$\mu$m and a width of 100\,nm. The colour wheel represents the spin direction which corresponds to a vector directed from the centre to the appropriate colour.}
\label{lowE}
\end{figure}

We start with the theoretical modelling of the expected spin structures. The results of the lowest energy simulations are presented in Figure~\ref{lowE}. The inset depicts simulations of the two main wall types in 100\,nm wide half rings, revealing a vortex domain wall for the thicker structure with $d=$ 20\,nm (top) and a transverse domain wall for the thinner structure with $d=$ 7\,nm (bottom). By comparing the calculated total energies for these two domain wall types for each ring size, the lowest energy state is extracted as represented in the phase diagram.             

The investigation of the stability of the different stable domain wall configurations in nanowires was first investigated by McMichael and Donahue via analytical modelling and micromagnetic simulation~\cite{McMichaelDonahue}. For the analytical calculation they assumed that the major contribution to the energy of the transverse domain wall was from the stray field energy due to the off-axis magnetization region, while the vortex domain wall is assumed to be dominated by the exchange energy contribution from the closely circulating magnetization around the core. By equating expressions for these two contributions, an analytical form of the width ($w$) vs. thickness ($d$) phase boundary was derived to be:
\begin{equation}w d = 16 \pi \ln\left( \frac{r_{\textrm{max}}}{r_\textrm{min}} \right) \frac{A}{\mu_0 M_{s}^{2}}\label{eqn:phase1},\end{equation} where $r_\textrm{max}$ represent the outer radius of the vortex and $r_\textrm{min}$ the radius of the vortex core, respectively, $\mu_0$ is the permeability of free space, $M_s$ is the saturation magnetization and A is the exchange constant with the definition of the exchange length as $\delta=(A/\mu_0 M_{s}^{2})^{1/2}$. Ignoring the weak logarithmic dependence, Eq.~\ref{eqn:phase1} is of the following form:

 \begin{equation}w \times d = C \times \delta^2 ,\label{eqn:phase}\end{equation} 

where for Permalloy the  constant, $C$, was determined from their simulations to be $128$~\cite{McMichaelDonahue}. For our results for iron the phase boundary has been fitted to this functional form, as shown in Figure~\ref{lowE}. The function can be seen to fit the phase boundary well and yields $w_{\textrm{crit}} \times d_{\textrm{crit}}=(756\pm17)$\,nm$^2$. If we calculate the expected value using the material parameters for iron of $A=2.1 \times 10^{-11}$\,J/s and $M_s=1.7 \times 10^{6}$\, A/m and the value of C from the work of McMichael and Donahue we get $w_{\textrm{crit}} \times d_{\textrm{crit}}=740$\,nm$^2$ which is in excellent agreement with our results. In comparison to other systems, the stability of the vortex domain wall is pushed to narrower and thinner structures than for either Py~\cite{thermalDW} or Co~\cite{cowallphase}, which can now be directly understood as arising from the materials properties dependence of Equation~\ref{eqn:phase} via the contribution from the exchange length: the increased saturation magnetization in Fe favours the low stray field energy vortex domain wall, so that it remains the lower energy wall type down to narrower and thinner structures compared to cobalt and Permalloy.

\begin{figure}[tb!]
\centering
\includegraphics[width=1\textwidth,keepaspectratio=true]{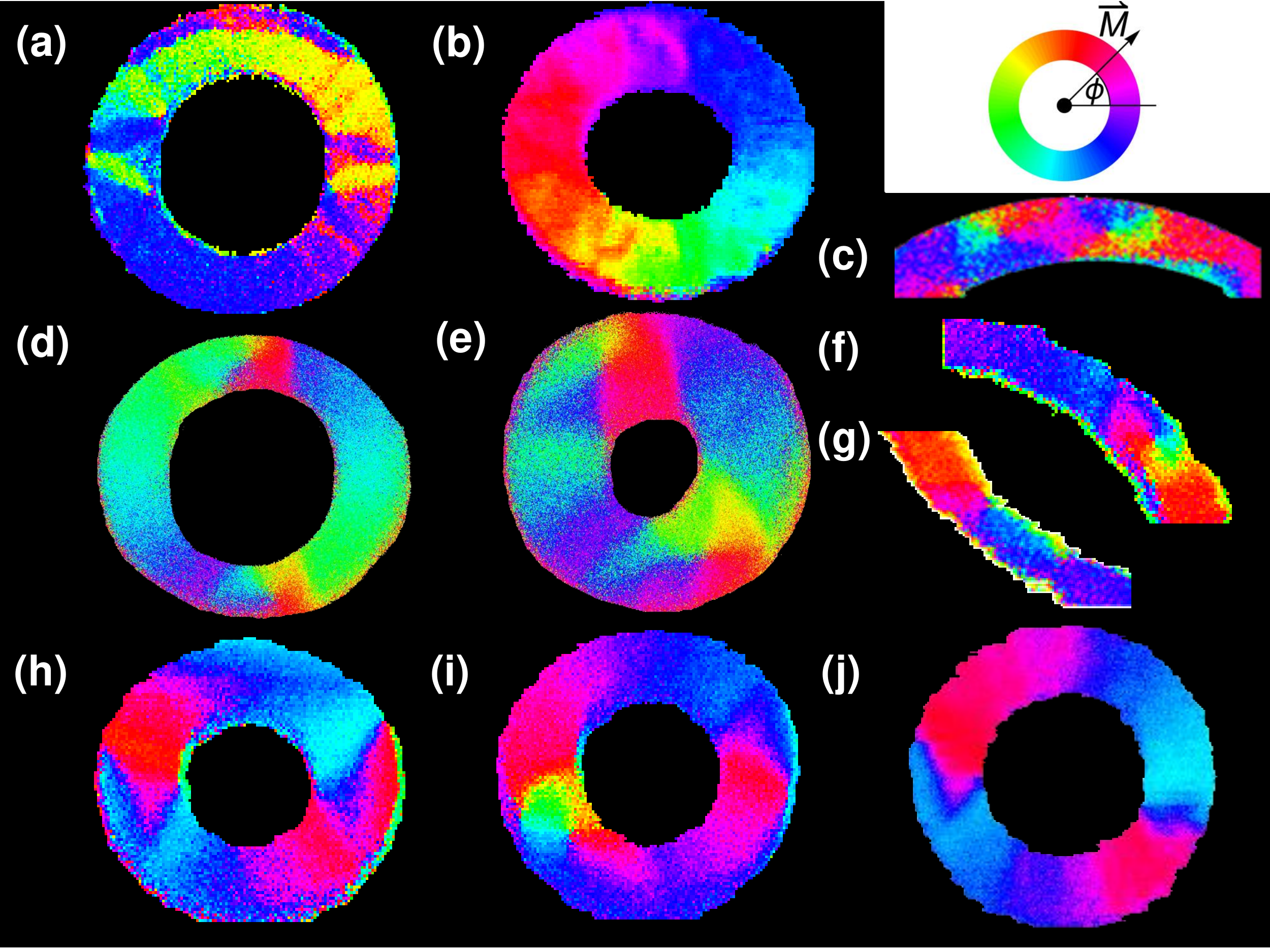}
\caption{Selected SEMPA images of domain wall configurations. Ring (a) has an O.D. of $3.5\,\mu$m. The rings (b)-(e) have  O.D.=\,$2\,\mu$m and all the remaining rings have an O.D.=\,$1\, \mu$m. The thicknesses of the rings are (a) 17\,nm, (b) 12\,nm, (c) 15\,nm, (d)/(e) 24.5\,nm, (f)-(i) 10\,nm and (j) are 5\,nm thick. The widths of the rings are (a) 750\,nm, (b) 550\,nm, (c) 300\,nm, (d) 400\,nm, (e) 650\,nm,  (f) 110\,nm,  (g) 90\,nm,  (h) 290\,nm, (i) 270\,nm and  (j) 230\,nm. The magnetic contrast direction is indicated by the colour wheel inset.}
\label{figureCombined}
\end{figure}

We then image the domain wall spin structures of different ring geometries. A selection of SEMPA images can be seen in Figure~\ref{figureCombined}. The SEMPA imaging simultaneously measures the two in-plane components of the magnetization which have been combined to give the full in-plane information of the magnetization vector as represented in the colour images here. The colour wheel inset represents the local spin direction, which corresponds to a vector directed from the centre of the wheel to the appropriate colour. Due to our initialization procedure, most rings were observed to be in the onion state and contain two domain walls, one head-to-head and one tail-to-tail, which are roughly aligned with the axis of the initializing field (along the x-direction for ring (a) and (f)-(j), but along the y-direction for the remaining rings). However the precise positioning of the walls is determined by local defects and hence the location of the walls in the experimental images is not always completely symmetric as would be expected for an ideal ring.

First we consider the larger structures as depicted in Figure~\ref{figureCombined} (a)-(e) where the dimensions of the rings are far away from the expected phase transition region. For the largest structure of 750\,nm width in (a) it can be seen that a fairly complicated domain structure is observed. In particularly wide annular structures it is expected that the simplest domain wall configurations are no longer the only accessible stable magnetic states and more complicated spin-structures have been predicted and observed experimentally due to the reduced shape anisotropy ~\cite{MKheadtohead, widerings, widehead}.  
In some structures ripple domains were observed as can be noticed in (b) which is also one of the few cases where annihilation of the two domain walls resulted in the vortex state with continually circulating magnetization around the ring and here the ripple domains are clearer to see. In the case of Permalloy, such ripple contrast is not found for these geometries due to the low intrinsic magnetocrystalline anisotropy, however, for Fe the magnetocrystalline anisotropy is significant albeit not as large as for Co. Since the structures are polycrystalline the effect of this anisotropy should cancel out over the whole structure, unlike with epitaxial samples~\cite{SEMPArings}, but at a local scale statistical variations in the anisotropy of individual grains lead to such characteristic contrast~\cite{ripple}.
Ring (c) shows an example of a more complicated double vortex structure in a 15\,nm thick ring which has previously been seen in Figure 6 (b) of ~\cite{MKheadtohead} for Py as a result of current induced wall transformations~\cite{currenttransformDW} and recently predicted as one of a variety of more complicated (meta)stable wall structures comprising multiple vortices and antivortices in wide strips~\cite{widehead}. Rings (d) \& (e) show vortex domain wall states in $24.5\,$nm thick rings with widths of 400 and 650\,nm, respectively, where the vortex wall can be seen to spread out in the wider rings due to reduced geometrical confinement. Overall the domain walls observed in this size range are of vortex type or are more complicated, which is consistent with the predictions of Figure~\ref{lowE}.

We now consider the domain walls observed in smaller and thinner rings, as represented by the second range of images (f)-(j) presented in Figure~\ref{figureCombined}. All of these rings show onion states with either vortex walls [(f) and left side of (i)] or transverse walls with varying symmetry. In the narrowest structures there is a tendency for symmetric transverse walls [e.g. (g)], while the transverse walls in the wider structures become very asymmetric [e.g. (h)]. However, it is immediately apparent that the phase boundaries are not completely well defined since in ring (i) both a vortex and asymmetric transverse wall are shown in the same structure and in ring (j) both a symmetric and asymmetric transverse wall are seen in the same structure.

\begin{figure}[htb]
\centering
\includegraphics[width=1\textwidth,keepaspectratio=true]{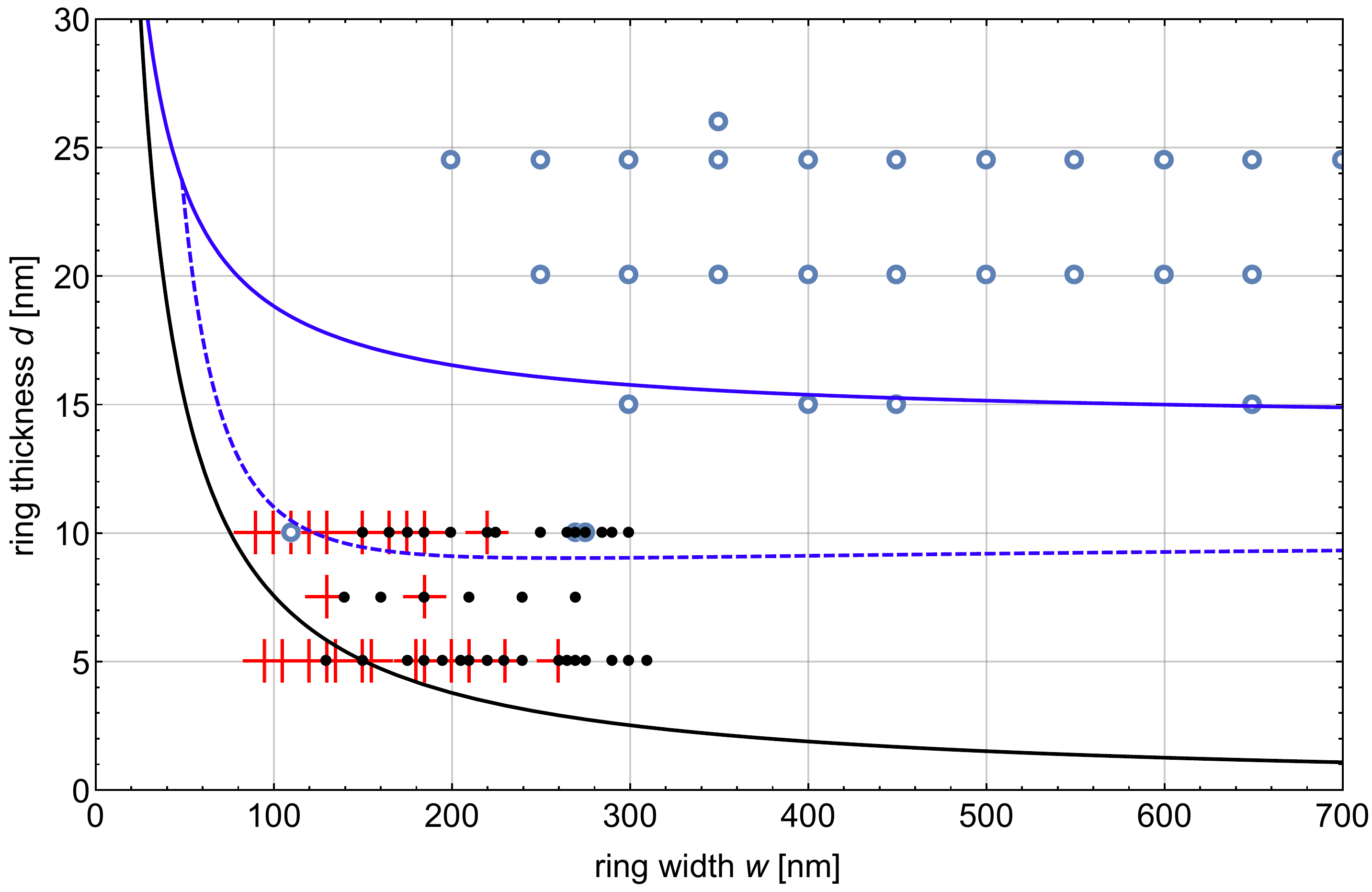}
\caption{Experimental phase diagram comparing the different observed spin structures and simulated phase boundaries. A red cross denotes symmetric transverse walls, a black dot asymmetric transverse walls and a blue circle vortex walls. The lowest energy transition is represented by the black solid line. The blue solid line represents the transition from a metastable transverse wall to a stable vortex domain wall by relaxing a field-saturated magnetic half ring, whereas the blue dashed line represents the transition from a symmetric to an asymmetric transverse wall under the same conditions.}
\label{thermal}
\end{figure}

Figure \ref{thermal} shows the distribution of all the observed domain wall types. The symmetric transverse walls are depicted as red crosses, asymmetric transverse walls as black dots and vortex walls as blue circles. The black curve represents the expected lowest energy phase transition as already presented in Figure~\ref{lowE}. Each studied Fe ring with a thickness of 15\,nm or more showed a vortex domain wall, whereas for structures with a thickness of 10\,nm only rings with widths of 110, 270 and 275\,nm exhibited vortex walls. The remaining structures at this and lower thicknesses showed either symmetric or asymmetric transverse walls. The distribution of these two domain wall types shows pronounced overlap, with both types seen in a range between 150 and 220\,nm for 10\,nm thick rings and a wider overlap region between 130 and 260\,nm for the thinnest 5\,nm thick rings. Furthermore, if we compare the observed types of domain wall to the predictions of the phase diagram in Figure~\ref{lowE}, serious quantitative discrepancies are apparent. Firstly, the observed asymmetric transverse walls are not covered at all by the analytical model and are not well represented by the simulations which calculated and compared the energies of just the two principal domain wall configurations which were set as the initial states. Secondly, while the observation of a vortex wall in Figure~\ref{figureCombined} (g) fits well with the lowest energy phase boundary, many of the observed transverse walls are at much larger widths and thicknesses than would be expected. The asymmetric wall type is not able to account for the discrepancy, since Nakatani et al. found that the asymmetric transverse wall phase belongs to an area below the vortex-transverse phase boundary and hence would not explain the occurance of transverse walls of either symmetric or asymmetric type in larger structures~\cite{refinedphasepy}.

\section{Discussion}
\label{sec:dis}
             
\begin{figure}[htb]
\centering
\includegraphics[width=0.75\textwidth,keepaspectratio=true]{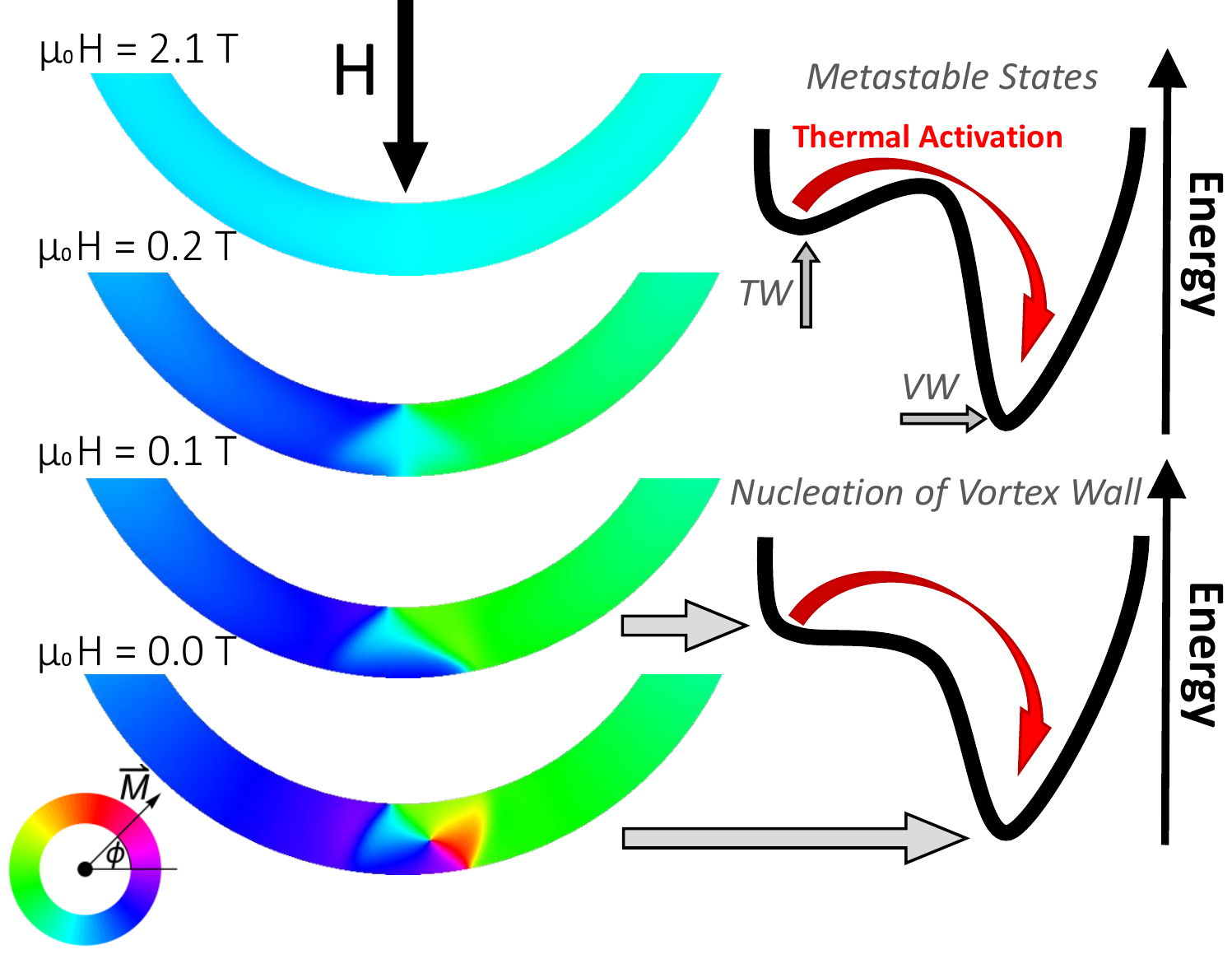}
\caption{The nucleation of a vortex domain wall in an Fe ring with a width of 100\,nm and a thickness of 15\,nm shows a transition through an asymmetric transverse wall at finite external magnetic field as shown in the 0K-simulations (left) and the lower right schematic diagram. For finite-temperature experiments thermal fluctuations occur which can cause a transition from a metastable transverse wall (TW) to an energetically favorable vortex wall (VW) as seen in the upper schematic diagram.}
\label{thermal0}
\end{figure}

\begin{figure}[tb!]
\centering
\includegraphics[width=0.7\textwidth,keepaspectratio=true]{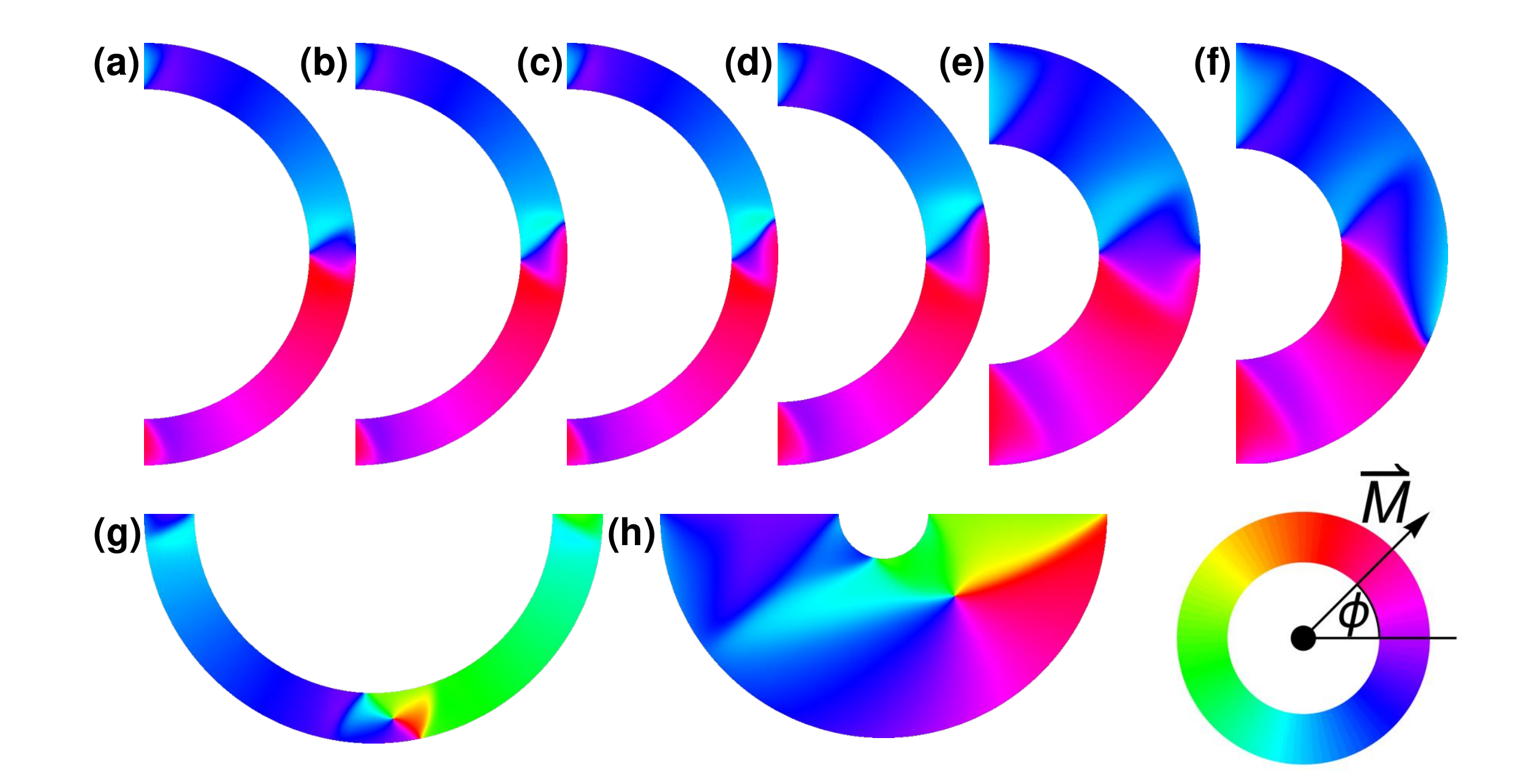}
\\\textbf{(i)}\\
\includegraphics[width=0.7\textwidth,keepaspectratio=true]{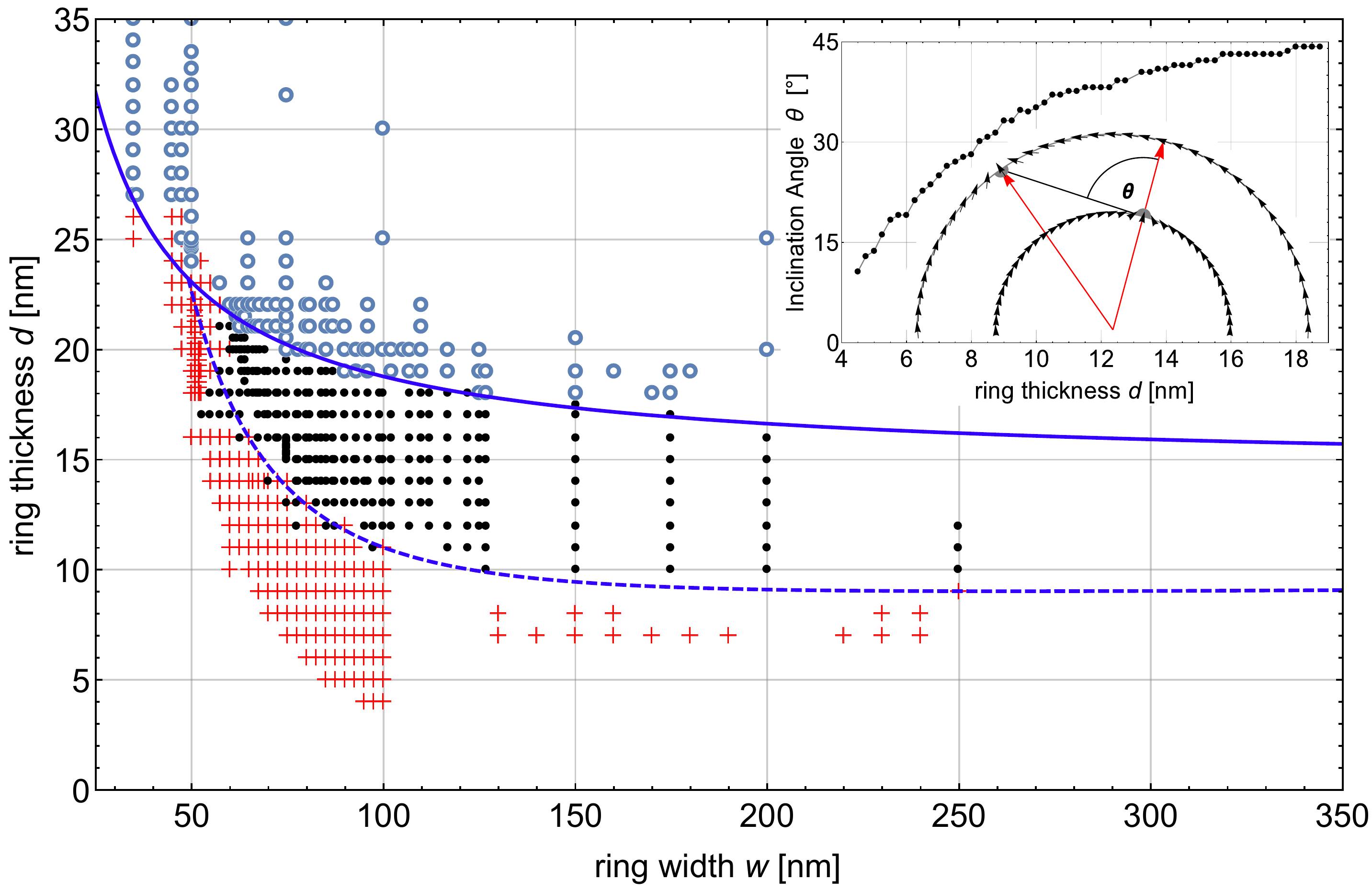}
\caption{The simulated spin structures obtained on relaxation of the magnetization from saturation, mimicking the experimental procedure. Three types of domain wall are observed, the vortex [(g) and (h)], symmetric transverse wall [(a) and (e)] and asymmetric transverse wall [(b)-(d) \& (f)] as depicted in the selected simulations. (a)-(c) \& (g) have width 110\,nm, (d) has width 150\,nm, (e)-(f) have widths of $240$/$250\,$nm and (h) is 400\,nm wide. The thicknesses are as follows: (e)-8\,nm, (a), (d) \& (f)-10\,nm, (h)-13\,nm, (b)-14\,nm, (c)-18\,nm and (g)-22\,nm. The graph in (i) shows the resulting domain wall phase diagram showing three clear regions of stability for the different domain wall types, evolving from symmetic transverse walls for smaller structures to vortex walls in larger structures, through an asymmetric transverse wall for certain intermediate dimensions. The inset displays the change in angle of the asymmetric transverse wall with the ring thickness for a fixed ring width of 100\,nm.}
\label{sims2}
\end{figure}

Whilst the simulated results provided a good fit to the analytical theory of McMichael and Donahue, the observation of the different types of transverse walls and the observation of transverse domain walls in the region of the phase diagram where the analytical model predicts that the vortex domain wall should be the lowest energy state indicates that there is more going on. Firstly, note that a direct comparison of the experimentally observed states with the analytical model requires that the occurring state is the global energy minimum of the system. However, in practice this would not always be expected to be the case. As mentioned above, the experimental domain wall states are initialized by relaxing the spin structures from saturation. Under these conditions, simulations show that the symmetric transverse wall forms initially~\cite{vortexheadtohead} and for a range of geometries this domain wall configuration can be a metastable state with an energy barrier that must be overcome in order to nucleate the vortex core as shown in Figure~\ref{thermal0}, leading to hysteretic switching between the two domain wall types with field~\cite{hystereticDW}. Secondly, the observation of asymmetric transverse walls in addition to symmetric ones demonstrates that the domain wall potential landscape is more complicated than just the simple picture of two stable spin structures, introduced above. In order to investigate these issues in more detail we performed a new set of simulations where we mimicked the experiment by saturating the half-ring along the symmetry axis with an external field of $\mu_0 H \approx$ 2.1\,T and gradually relaxed the field in logarithmic steps to reveal the stable zero field states. The results of this second set of simulations are presented in Figure~\ref{sims2}. Pictures (a)-(h) display typical results across a range of geometries with the wall orientation and colour-code adjusted to enable ease of comparison with the experimental results. Firstly, for the smallest structures such as the 10\,nm thick 110\,nm wide ring in (a) we see a symmetric transverse wall. On increasing the thickness [(b), (c)] or width [(d), (f)], this then transforms into an asymmetric transverse wall with gradually increased asymmetry. For thinner structures [(e)], meanwhile, the symmetric transverse wall is stable up to higher widths. Finally for the thickest [(g)] and widest [(h)] structures the vortex domain wall emerges. This trend and the form of the spin structures show excellent agreement with the structures observed experimentally. This is clear on comparing e.g. Figure~\ref{sims2} (e) with the right side of Figure~\ref{figureCombined} (j) for the symmetric transverse wall, Figure~\ref{sims2} (d)/(f) with Figure~\ref{figureCombined} (h) and the right of (i) for the asymmetric transverse walls and Figure~\ref{sims2} (g)/(h) with Figure~\ref{figureCombined} (d)/(e) for the vortex walls. 

For a more quantitative treatment the observed domain wall types for all simulations are presented in the graph of Figure~\ref{sims2} (i). When comparing with the previous simulations the first point to note is that now there are two phase boundaries; a principal upper boundary which separates the vortex domain walls from the transverse walls, as before, and a second sub-phase boundary which separates the symmetric transverse walls from the asymmetric ones. The main phase boundary now corresponds to much larger $w\times d$ than before, resulting in a larger range of geometries where the transverse domain walls occur, reflecting the incorporation of states where this spin structure is a metastable state. The new sub-phase boundary is similar in shape to that presented for Py in~\cite{refinedphasepy}, however, it should be noted that the phase boundary presented in that reference represents the lowest energy configurations on comparing the energies of the three domain wall types. For our Fe phase diagram, which mimics the experiment, we find a larger region where the asymmetric transverse wall is expected and find the asymmetric walls for narrower rings down to $w=50$\,nm as compared to $w=150$\,nm in the previous case of Py. For the newly observed asymmetric transverse domain wall region, the inset of Figure~\ref{sims2} (i) plots the evolution in tilting angle of the asymmetric domain wall with thickness for a fixed ring width of 100\,nm. Here, the inclination angle of the wall is defined as depicted on the plot. The angle is observed to undergo a monotonic increase with increasing thickness of the ring until it approaches 45$^\circ$, corresponding to magnetization aligned with the edge in these particular structures although in wider structures larger angles are observed. This behaviour can be understood from the different thickness dependencies of the contributing energy terms. The increased asymmetry of the wall reduces the component of magnetization directed normal to the wire edge and hence reduces the stray field, at the expense of an increase in exchange energy. In this sense, the increasing asymmetry of the wall is a precursor to the formation of the full vortex wall. Since the exchange energy is linear in thickness while the magnetostatic energy is quadratic, the energy gain in reducing the stray field becomes more significant with increasing film thickness, resulting in the observed behaviour.
To quantitatively compare the experimental results with the collected simulation results, Figure~\ref{thermal} also includes the phase boundaries from this second set of simulations as represented by the blue lines. What is apparent is that both vortex and transverse domain wall are still observed in the region between the principal solid-line boundaries from the two calculation methods which means that \emph{neither} boundary is a good fit to the experiment. However, one of the remaining factors that we need to consider is the influence of thermal effects, since the simulations only provide the 0\,K states, while the experiments are performed at ambient conditions of 293\,K. While the transverse wall initially forms on relaxation from saturation, it is only a metastable state in the region between the two principal phase boundaries and hence thermal activation is able to transform the wall to the vortex configuration which is the lowest energy state, as depicted schematically in Figure~\ref{thermal0}. Whether this occurs experimentally depends on the measurement temperature and the height of the energy barrier, which in the absence of pinning is determined by the extent to which the geometrical parameters differ from those of the upper phase boundary. Taking all the experimental and theoretical data together we come to the key result that we can divide the phase diagram into three main regions. Firstly, below the lower principal phase boundary, is a region where transverse domain walls should always be observed in the experiment. Secondly, in the region between the two main boundaries represented by the solid curves, we have a region where either type of wall may be observed, depending on the temperature. As the temperature is increased the effective experimental transition between wall types is expected to move from the upper boundary to the lower one, although due to thermal fluctuations this is not expected to be a rigid boundary at any given temperature and in the vicinity of this effective boundary both types of wall may be observed for the same geometry. Finally, above the upper boundary, is a third region where vortex or more complicated domain walls would always be expected. This is in agreement with the experimental observations. For our room temperature measurements we observe two vortex domain walls for the 10\,nm thick samples at similar widths of 270 and 275\,nm, suggesting that the room temperature transition is close to this point. This is also consistent with the fact that we did not observe vortex walls in the thinner samples since this vortex wall would be expected to be stable only for larger widths than for the thicker samples (for such geometries no experimental data were recorded due to the stray field considerations outlined above). We note that whilst we also observe a vortex domain wall for the 110\,nm wide, 10\,nm thick ring, this is at a much lower width than the other experimentally observed vortex states, suggesting that thermal activation is unlikely to be the only influence in this particular case, as elaborated below.

The final ingredient that needs to be taken into consideration in the experimental work, that is not present in the simulations, is the effect of magnetization pinning due to defects, edge roughness and the local influence of magnetocrystalline anisotropy from the small crystallites that compose the polycrystalline layers. Indeed, the latter effect is already evidenced in some of the images due to the emergence of ripple domain contrast, as mentioned above. Such pinning effects can also lead to the stabilization of higher energy metastable spin configurations~\cite{multiplicity}. 
To get more insight into the influence of defects we have analyzed a ring structure with a particularly strong edge roughness, which we can gauge from the images of the sample topology which are automatically acquired during the SEMPA imaging (not shown). The domain configurations for opposite sides of this same structure are depicted in Figure~\ref{figureCombined} (f) \& (g), with a vortex wall seen on the wider side of the ring but a transverse wall on the narrower side. Whilst this vortex wall is not inconsistent with the lowest energy calculations, as mentioned above, it is far from the other regions of observed vortex walls. The roughness induced variations in wire width, however, can significantly affect the domain wall energy potential landscapes as seen for artificially defined domain wall traps~\cite{holoDW, asymTDWdepin}, thereby influencing the observed wall type. Alternatively material defects or a local change of the saturation magnetization~\cite{hashim}, for example, could help to promote the vortex wall by acting as preferential nucleation sites for the vortex core in such a structure. 
The effect of defects varying in kind, strength and position, would also explain the nature and position of the observed experimental phase boundaries. In the previously determined phase diagram for Py the experimental room temperature phase boundary was found to be close to the predictions of the analytical model, indicating that thermal activation is able to have a large effect in that system~\cite{thermalDW}. Here, however, the experimental phase boundary seems to be in the middle of the region of metastability which can be attributed to the increased pinning in the polycrystalline Fe system showing that the materials properties influence this very strongly. An even more extreme effect of pinning was observed in the Co system where the experimental phase boundary was found to be very close to the upper 0\,K limit ~\cite{cowallphase} and more generally it would be expected that it is not only the material that will affect the location of the observed boundary, by way of the intrinsic magnetocrystalline anisotropy, but also the particular growth conditions. Finally, the presence of defects can also explain the broadness of the experimentally observed phase boundary between symmetric and asymmetric transverse walls. Such defects are likely to promote asymmetry in the walls, stabilizing the asymmetric configuration over a larger range of geometries than in the simulations of defect free systems and leading to a range of widths where, depending on the individual structure, either an asymmetric or symmetric transverse wall may form.

\section{Conclusion}
\label{sec:con}

In conclusion, we provide a comprehensive investigation of the complex phase boundaries of different domain wall types by varying the energetic contributions to the stable domain wall spin structures in mesoscopic Fe rings via micromagnetic simulations and high resolution magnetic imaging using SEMPA. The lowest energy domain wall states are well described by the previously developed analytical model of McMichael and Donahue taking into account the competition between exchange and dipolar energy. In order to mimic the experiment, simulations are performed that relax the magnetization states from saturation, revealing in addition to the frequently studied vortex domain wall and symmetric transverse wall, regions of stability of asymmetric transverse domain walls under experimentally relevant initialization conditions. 
Here the general trend from symmetric transverse domain wall for thin and narrow rings, to an asymmetric transverse domain wall with gradually increasing tilt angle for thicker and wider structures and finally to a vortex domain wall for the thickest and widest rings, can again be understood as arising from the competition between exchange and dipolar energy contributions which evolve with the geometry. Since domain walls of different type and degrees of asymmetry often display very different dynamic behaviours, it is crucial to understand how different factors can be used to tailor the spin structure and through this the properties of a device. As this present study shows, the experimentally observed domain wall configurations are actually the result of a complex interplay of several factors which all need to be considered. Due to the attractive magnetic properties of iron including a relatively large spin polarization and a lower magnetocrystalline anisotropy than Co, as well as a simple growth procedure, the results presented here show promise for using Fe in devices based on domain wall motion. In particular, for robust device operation it is necessary to find regions of the phase diagram where little variance in the domain wall type is observed and the diffuse boundary regions identified here, for example, would be detrimental for reliable device performance. Whilst vortex walls are stable for a wide range of geometries at particularly large ring widths and thicknesses, the symmetric transverse walls are only reproducibly seen in very narrow and/or thin structures.

\section*{Acknowledgments}

We acknowledge financial support from the via the DFG collaborative research centre SFB/TRR 173 SPIN+X, the Graduate School of Excellence Materials Science in Mainz (MAINZ) GSC 266, the Carl-Zeiss-Foundation, FP7-PEOPLE-2013-ITN WALL (Grants No. 608031) and ERC-PoC-2014 MultiRev (Grants No. 665672).

\bibliography{Domain_Wall_Spin_Structures_in_Mesoscopic_Fe_Rings_probed_by_High_Resolution_SEMPA}
\bibliographystyle{apsrev4-1}

\end{document}